# Positive-unlabeled convolutional neural networks for particle picking in cryo-electron micrographs

Tristan Bepler[1,2], Andrew Morin[2,3], Julia Brasch[4], Lawrence Shapiro[4], Alex J. Noble[5,*],

and Bonnie Berger[2,3,**]

[1] Computational and Systems Biology, MIT, Cambridge, MA, USA

[2] Computer Science and Artificial Intelligence Laboratory, MIT, Cambridge, MA, USA

[3] Department of Mathematics, MIT, Cambridge, MA, USA

[4] Department of Biochemistry and Molecular Biophysics, Mortimer B. Zuckerman Mind Brain Behavior Institute, Columbia University, NY, NY, USA

[5] National Resource for Automated Molecular Microscopy, Simons Electron Microscopy Center, New York Structural Biology Center, NY, NY, USA

* Corresponding author for cryoEM experiments: anoble@nysbc.org

** Corresponding author: bab@mit.edu

**Abstract**

Cryo-electron microscopy (cryoEM) is an increasingly popular method for protein structure determination. However, identifying a sufficient number of particles for analysis (often >100,000) can take months of manual effort. Current computational approaches are limited by high false positive rates and require significant ad-hoc post-processing, especially for unusually shaped particles. To address this shortcoming, we develop Topaz, an efficient and accurate particle picking pipeline using neural networks trained with few labeled particles by newly leveraging the remaining unlabeled particles through the framework of positive-unlabeled (PU) learning. Remarkably, despite using minimal labeled particles, Topaz allows us to improve reconstruction resolution by up to 0.15 Å over published particles on three public cryoEM datasets without any post-processing. Furthermore, we show that our novel generalized-expectation criteria approach to PU learning outperforms existing general PU learning approaches when applied to particle detection, especially for challenging datasets of non-globular proteins. We expect Topaz to be an essential component of cryoEM analysis.

**Introduction**

Transmission electron microscopy (TEM) of purified proteins vitrified on thin metal EM grids, called single particle cryo-electron microscopy (cryoEM), is an emerging technology capable of resolving high resolution structures of proteins in near-native states. CryoEM projection images, or micrographs, can contain hundreds or thousands of individual protein projections, called particles. By identifying a sufficient number of particles in a sufficient number of orientations, a 3D reconstruction of the purified protein can be solved by aligning the particles in 3D Fourier space[1]. However, due to the low signal-to-noise ratio of cryoEM images, large numbers of observations are required for accurate reconstruction. Studies have shown a monotonically-increasing linear dependence between the logarithm of the number of particles included in a single particle alignment and the inverse of the resolution of the resulting 3D reconstruction[2]. The concentration of protein on EM grids, the efficiency of data collection, and the completeness and accuracy of particle identification are all factors affecting the total number of particles that can be used for downstream reconstruction and hence the achievable resolution. In particular, accurate particle identification, also known as particle picking, is a major bottleneck, often taking weeks or even months with current workflows for small or non-globular particles, due to variability in particle shapes and structured noise in cryoEM micrographs.

A variety of methods have been developed for automation of particle picking. The most common are Difference of Gaussians (DoG) and template-based approaches[3–7]. DoG is a detection method limited to spherical, compact particles of known size since it matches these features directly. Template-based approaches, on the other hand, search micrographs for regions matching available particle templates based on cross correlation. These templates can be derived from known structures or using hand-labeled particles. Template derivation from known structures, in particular, can produce large numbers of false positives or fail for atypical particles[8–10].

Recently, to address the shortcomings of these techniques, new methods based on convolutional neural networks (CNNs) have been proposed[11–13]. These methods use positive and negative labeled micrograph regions to train CNN classifiers which are used to predict labels for the remaining regions. However, due to factors like low signal-to-noise ratio, structured background, and the distribution of particle morphologies, researchers must label a large number of regions for training — a non-trivial and time-consuming task. Moreover, the diverse characteristics of negative data make it difficult to manually label a representative set of negative examples, and hence the number of labeled negatives must be an order of magnitude larger than the number of positives to achieve acceptable performance[14]. Thus, these methods have seen only limited adoption by the cryoEM community and hand-labeling remains the gold standard.

To overcome the challenges inherent in current automatic particle picking methods, we newly frame the problem of particle-picking as a positive-unlabeled (PU) learning problem in which we seek to learn a classifier using a small number of labeled positive regions and the remaining unlabeled micrograph regions. PU learning has proved to be an effective paradigm when working with partially labeled data, primarily with application to document classification[15], time series classification[16], and anomaly detection[17]. Recent work in machine learning has generally explored PU learning for neural network models[18] based on estimating the true positive-negative risk, but overfitting remains a challenging problem for PU learning. In order to address this, we approach PU learning as a constrained optimization problem in which we wish to find classifier parameters to minimize classification errors on the labeled data subject to a constraint on the expectation over the unlabeled data. By imposing this constraint softly with a novel generalized expectation (GE) criteria, we are able to mitigate overfitting and train high accuracy particle classifiers using very few labeled data points. Furthermore, by combining our PU learning method with autoencoder-based regularization, we can further reduce the amount of labeled data required for high performance.

Here, we present Topaz, a pipeline for particle picking using convolutional neural networks with PU learning. We demonstrate that by using Topaz with only 1,000 labeled examples, we are able to improve the 3D structure resolution of several publicly available cryoEM datasets by up to 0.15 Å over the published particle sets, which were constructed based on significant manual curation efforts. Remarkably, this improvement is without any ad hoc post-processing typically required for high resolution structures; we feed Topaz predictions directly into alignment and reconstruction. We also find that Topaz achieves a remarkably low false positive rate even when using a relaxed predicted probability threshold to retrieve large particle sets. Finally, we compare our GE based PU learning techniques with other recent PU learning methods and demonstrate that our PU learning approach enables particle picking with minimal labeled examples, even on a challenging dataset containing stick-like particles with low signal-to-noise.

Our source code is freely available for academic (https://github.com/tbepler/topaz) and the program runs efficiently on a single GPU computer. Topaz is currently being integrated into Appion[20] and may be integrated into other cryoEM software suites.

# Results

## 1. The Topaz Pipeline

The Topaz particle picking pipeline is composed of three main steps (Figure 2): (1) whole micrograph preprocessing with a mixture model newly designed to capture micrograph statistics, (2) neural network classifier training with our PU learning framework, and (3) micrograph segmentation and particle coordinate extraction by non-maximum suppression. Our implementation of this pipeline is freely available for academic use at https://github.com/tbepler/topaz.

*Micrograph preprocessing*

Prior to classifier training, micrographs must first be downsampled and normalized. Downsampling is critical for reducing pixel level noise and making training and particle extraction more efficient. Micrographs are then normalized using a per-micrograph scaled two-component Gaussian mixture model (see Methods). This captures and corrects for large-scale intensity differences between micrographs and the bi-modality of pixel values caused by micrographs containing dark grid sections.

*Classifier training from positive and unlabeled data*

The core advancement of Topaz is our ability to leverage unlabeled data when training particle classifiers. We frame particle picking as a PU learning problem in which we seek to learn a classifier that discriminates between particle and non-particle micrograph regions given a small number of labeled particles and many unlabeled micrograph regions. CNN classifiers are trained using minibatched stochastic gradient descent with a novel objective function, GE-binomial (see

Methods for details), which explicitly models the sampling statistics of minibatch training to regularize the classifier's posterior over the unlabeled data. Combining this with an optional autoencoder module allows high-accuracy classifiers to be trained despite using very few positive examples (Methods). This approach allows us to overcome overfitting problems associated with recent PU learning methods developed for neural networks in domains other than cryoEM analysis and to effectively pick particles in challenging cryoEM datasets as we demonstrate in our experiments.

*Micrograph segmentation and particle extraction*

Given a trained CNN particle classifier, we extract predicted particle coordinates and their associated predicted probabilities. First, we calculate the per pixel predicted probabilities by convolving the classifier over each micrograph. Then, to extract coordinates from these dense predictions, we use the well known non-maximum suppression algorithm to greedily select high scoring pixels and remove their neighbors from consideration as particle centers. This yields a list of predicted particle coordinates and their associated model scores for each micrograph.

**2. Topaz improves structure resolutions with no postprocessing**

Because the end goal of single-particle cryoEM is to produce a high resolution 3D structure, we evaluate the full Topaz particle picking pipeline by utilizing picked particles in reconstructions for three cryoEM datasets containing T20S proteasome (EMPIAR-10025), 80S ribosome (EMPIAR-10028), and rabbit muscle aldolase (NYSBC-aldo). Each of these datasets already has a curated set of particles giving high quality reconstructions which we compare with particles predicted by Topaz trained with 1,000 positives based on reconstruction quality (Methods). We standardize the reconstruction procedure by using cryoSPARC homogeneous refinement on the raw Topaz particle sets (i.e. no postprocessing was applied) and published particle sets with identical settings for each dataset (Methods). By considering the reconstruction resolution at decreasing probability thresholds (increasing numbers of particles) predicted by Topaz, we select the particle set that optimizes the resolution for each dataset. We find that Topaz is able to retrieve significantly more good particles than were present in the curated particle sets, finding 3.22, 1.72, and 3.68 times more particles in EMPIAR-10025, EMPIAR-10028, and NYSBC-aldo respectively. Remarkably, this led to an improvement in the reconstruction resolution by ~0.15 Å for EMPIAR-10025 and ~0.05 Å for EMPIAR-10028 (Figure 3) despite performing no postprocessing (i.e. particle filtering with 2D or 3D class averaging or iterative reconstructions removing particles poorly fitting the 3D model) on the Topaz particle sets. All predicted particle coordinates were fed directly into the reconstruction pipeline. Furthermore, the 3.0 Å structure we report for EMPIAR-10028 is, to our knowledge, the highest resolution structure ever reported for this dataset. We improve the resolution by 0.1 Å over the best structure reported in the EM map challenge[21] simply by picking particles with Topaz trained using 1,000 initial examples. For NYSBC-aldo, although Topaz finds many more particles than were in the published dataset, both particle sets achieve the same reconstruction resolution (2.63 Å by $FSC_{0.143}$), suggesting that the ~200k particles in the published set is already sufficient to reach the resolution limit of the data given standard reconstuction methods. We verify that the additional particles found by Topaz are good particles by performing reconstructions using only the newly picked particles (i.e. we remove the entire published particle set from the Topaz particle set and perform reconstruction) and we find nearly identical structures (Figure 3).

**3. Topaz particle predictions are well-ranked and contain few false positives**

We next quantify the quality of the particles predicted by Topaz over varying predicted probability thresholds by calculating the reconstruction resolution and estimating the number of false positive particles based on 2D class averaging. For each dataset, reconstructions are calculated using particles predicted by Topaz at decreasing probability cutoffs (Figure 4a). We find that the resolution of Topaz structures increases as we include more good particles and then drops once the threshold becomes small and too many false positives are included as demonstrated by the dip in resolution for the last threshold of EMPIAR-10025. Furthermore, we compare these curves with those obtained by randomly subsampling the published particle sets and find that Topaz particles quickly match the resolution of the published particles for the proteasome and ribosome datasets. For the aldolase dataset, we see that more Topaz particles are required to match and then exceed the resolution of the curated particle set. This could be because Topaz does not find enough side views of the particle until the probability is sufficiently lowered whereas the curated dataset has been filtered to be enriched for these views (Supplementary Figure 5).

We also classified the particle sets at each threshold into ten classes and manually examined the class averages to determine whether each class represented true particles or false positives. As expected, we find that as the probability threshold is decreased, the fraction of false positives increases (Figure 4b). However, the number of false positives remains remarkably low even at relaxed thresholds. Furthermore, particles appear to be well-ranked in that noisy or unusual particle classes only start to appear at low thresholds. For example, the T20S proteasome dataset is contaminated with gold particles which appear as dark spots in the micrographs. Particles in close proximity to gold are only selected as the probability threshold is decreased (Figure 4c). Similar trends can be observed in the ribosome (Supplementary Figure 4) and aldolase (Supplementary Figure 5) class averages.

**4. Our GE criteria based PU learning framework outperforms other PU learning methods on cryoEM datasets**

*Comparison of PU learning methods*

We consider two generalized expectation-based approaches to PU learning. Given the positive class prior, $\pi$, we constrain the classifier using the unlabeled data by matching the classifier's expectation to $\pi$ using the KL-divergence (GE-KL) or using the cross entropy between the classifier's posterior and the Binomial distribution prior over the number of positives in each minibatch (GE-binomial) (Methods). To evaluate the effectiveness of our GE-based PU learning methods, we benchmark against the recent non-negative risk estimator approach of Kiryu et al.[18] (NNPU) and the naive approach in which unlabeled data are considered as negative for classifier training (PN) on two additional cryoEM datasets. This is important to keep our PU learning methods development separate from the full Topaz evaluation above. The first dataset, EMPIAR-10096, is a publicly available dataset containing influenza hemagglutinin trimer particles and the second, Shapiro-lab, is a challenging dataset provided by the Shapiro lab containing a stick-like particle with low signal-to-noise. For purposes of comparison, we simulated positively labeled datasets of varying sizes by randomly subsampling the set of the all positive examples within the training set of each dataset.

We find that across all experiments, classifiers trained with our GE criteria-based objective functions dramatically outperform those trained with the NNPU or PN methods. Generally, GE-binomial and GE-KL classifiers display similar performance with a few important exceptions where GE-binomial gives better results. For the dataset with more compact particles, EMPIAR-10096, GE-binomial gives significantly ($p<0.05$ by Student's paired t-test) better test set average-precision scores than GE-KL when the number of data points is tiny (10 positive examples; Figure 5a). At larger numbers of positives, both methods are statistically equivalent. On the challenging Shapiro-lab dataset, GE-binomial significantly outperforms GE-KL at 1,000 labeled examples ($p<0.05$) whereas GE-KL gives better results ($p<0.05$) within the 50-250 range of labeled examples. These results indicate that our GE based PU learning approaches dramatically outperform previous PU learning methods, enabling particle picking despite few labeled positives on the challenging Shapiro-lab dataset and substantially improving picking quality on the easier EMPIAR-10096. Although GE-binomial and GE-KL perform similarly in this experiment, we do find that GE-binomial outperforms GE-KL in the two important cases of 10 easy particles and 1,000 difficult particles.

*Augmentation with autoencoder*

We next consider whether classifier performance can be improved when few labeled data points are available by introducing a generator network with corresponding reconstruction error term in the objective to form a hybrid classifier+autoencoder network (Methods). We hypothesized that including this reconstruction component would improve the generalizability of the classifier when few labeled data points are available by requiring that the feature vectors given by the encoder network be descriptive of the input – acting as a sort of machine learning technique known as regularization.

We test this by training classifiers with different settings of the autoencoder weight, $\gamma$, and varying numbers of labeled data points, $N$, on the EMPIAR-10096 and Shapiro-lab datasets. We compare models trained with $\gamma = 0$ (no autoencoder), $\gamma = 1$, and $\gamma = 10/N$. For each setting of $\gamma$ and $N$, we train 10 models with different sets of $N$ randomly sampled positives and calculate the average-precision score for each model on the test split of each dataset. We find that including the decoder network with reconstruction error term in the objective ($\gamma = 1$ and $\gamma = 10/N$) improves classifier performance in the few labeled data points regime (Figure 5b). As the number of data points increases, the benefit of using the autoencoder decreases and then hurts classifier performance due to over-regularization. Our results from both datasets suggest that using the autoencoder with $\gamma = 10/N$ gives best results when $N \leq 250$ and that not using the autoencoder is best for $N > 250$. Combined with PU learning, autoencoder-based regularization is an effective method to further improve classifier performance when few labeled positives are available.

## Discussion

CryoEM is revolutionizing structural biology with widespread applications ranging from basic biology to the understanding of disease-linked proteins and the development of novel therapeutics. Fully realizing the promise of cryoEM and achieving rapid turnaround from imaging to structure determination requires state-of-the-art computational methods. To this end, we present Topaz, a particle picking pipeline using PU learning – a framework naturally suited to the task of particle picking where only a small number of labeled positive examples are available for training. We show empirically that our GE-criteria-based approach outperforms other PU learning techniques for particle detection. In addition, augmenting the neural network classifier with an autoencoder further improves performance in the regime of very few labeled data points (n < 500). Topaz enables particle picking for unusually shaped particles with low signal-to-noise and, remarkably, allows us to improve structure resolution without any time consuming and ad-hoc post-processing.

Although we use a simple CNN architecture with reasonable default hyperparameters and show that it performs well on these datasets, any model architecture that can be trained with gradient descent can use our GE-criteria objective functions to learn from positive and unlabeled data. Furthermore, additional hyperparameter tuning, such as L2 or dropout regularization, can improve model performance. The only hyperparameter introduced by our objective function, and other positive-unlabeled objectives that we consider, is the unknown positive class prior. Although this parameter should also be chosen by cross validation, we observed that our results were relatively insensitive to its choice (Supplemental Figure 3). Our novel GE-binomial PU learning method could also have widespread utility for object detection in other domains, for example in light microscopy or medical imaging where positive labels are frequently incomplete.

Additionally, although we proposed GE-binomial for positive-unlabeled learning, it is straightforward to extend to the full semi-supervised case (where some labeled negative regions are provided) by taking the expectation of the loss over all labeled data in the first term.

By reporting a predicted probability of being a particle in addition to particle coordinates, Topaz allows researchers to take particle sets of varying size by choosing a probability threshold at which to select particles. This allows particles to be included iteratively by lowering the threshold until reconstruction resolution stops improving. However, it is also possible for reconstruction algorithms to explicitly take these probabilities into account when determining 3D structures.

Topaz requires researchers to label far fewer particles to achieve high quality predictions. It performs well independently of particle shape, opening automated picking to a wide selection of proteins previously too difficult to locate computationally. In addition, our pipeline is computationally efficient – training in a few hours on a single GPU and producing predictions for hundreds of micrographs in only minutes. Furthermore, once a model is trained for a specific particle, it can be applied to new imaging runs of the same particle. Topaz greatly expedites structure determination by cryoEM, enabling particle picking for previously difficult datasets and reducing the manual effort required to achieve high resolution structures.

# Methods

## 1. Dataset description

Aligned and summed micrographs and star files containing published particle sets were retrieved from EMPIAR for datasets EMPIAR-10025, EMPIAR-10028, and EMPIAR-10096. Aligned and summed micrographs and hand labeled particle coordinates were provided by the Shapiro lab for the Shapiro-lab dataset. Aligned and summed micrographs and a curated in-house particle set were provided by the New York Structural Biology Center for the NYSBC-aldo dataset. Micrographs for each dataset were downsampled to the resolution specified in table 1 and normalized as described in the following section. Each dataset was then split into training and test sets at the micrograph level. The number of micrographs and labeled particles in each split are also reported in table 1.

## 2. Micrograph normalization

Images were then normalized using a per-image scaled two component Gaussian mixture model. Given $K$ images, each pixel is modeled as being drawn from a two component Gaussian mixture model, parameterized by $\pi$, the mixing parameter, $\mu_0, \sigma_0, \mu_1$, and $\sigma_1$, the means and standard deviations of the Gaussian distributions, with a scalar multiplier for each image, $\alpha_{1...K}$. Let $x_{i,j,k}$ be the value of the pixel at position $i,j$ in image $k$, it is distributed according to

$$z_{i,j,k} \sim Bernoulli(\pi)$$

$$x_{i,j,k} \mid z_{i,j,k} \sim Gaussian\left(\alpha_k \mu_{z_{i,j,k}}, \left(\alpha_k \sigma_{z_{i,j,k}}\right)^2\right)$$

where $z_{i,j,k}$ is a random variable denoting the component membership of the pixel. The maximum likelihood values of the parameters $\pi, \mu_o, \mu_1, \sigma_0, \sigma_1$ and $\alpha_{1...K}$ are found by expectation-maximization for each data set. Then, the pixels are normalized by first dividing by the image scaling factor and then standardizing to the dominant mixture component. Let $\mu', \sigma'$ be $\mu_0, \sigma_0$ if $\pi < 0.5$ and $\mu_1, \sigma_1$ otherwise, then the normalized pixel values $x'_{i,j,k}$ are given by

$$x'_{i,j,k} = \left(\frac{x_{i,j,k}}{\alpha_k} - \mu'\right) / \sigma'$$

## 3. PU learning baselines

Let P be the set of labeled positive micrograph regions (centered on a particle), and U be the set of unlabeled micrograph regions where $\pi$ is the fraction of positive examples within U. Then, the task is to learn a classifier ($g$) that discriminates between positive and negative regions given P and U. When $\pi$ is small, treating the unlabeled examples as negatives for the purposes of classifier training with the following standard loss minimization objective can be effective

$$\pi E_{x \sim P}[L(g(x), 1)] + (1 - \pi) E_{x \sim U}[L(g(x), 0)] \qquad \text{(PN)}$$

However, in general, this approach suffers from overfitting due to poor specification of the classification objective - it is minimized when positives are perfectly separated from unlabeled data points. To address this, Kiryo et al.[18] recently proposed an unbiased estimator of the true positive-negative classification objective for positive and unlabeled data with known $\pi$ and a non-negative estimator (PU) which is shown to reduce overfitting still present in the unbiased estimator.

## 4. PU learning with generalized expectation criteria

Here, we adopt an alternative approach to positive-unlabeled learning not based on estimating the PN misclassification risk. Instead, we observe that the unlabeled data with known $\pi$ can be used to constrain a classifier such that it minimizes the classification loss on the labeled data and matches the expectation ($\pi$) over the unlabeled data. In other words, we wish to find the classifier, g, that minimizes $E_{x\sim P}[L(g(x),1)]$ subject to the constraint $E_{x\sim U}[g(x)] = \pi$. This constraint can be imposed "softly" through a regularization term in the objective function with weight $\lambda$:

$$E_{x\sim P}[L(g(x),1)] + \lambda KL(E_{x\sim U}[g(x)] \parallel \pi) \qquad \text{(GE-KL)}$$

In this objective function, we impose the constraint through the KL-divergence between the expectation of the classifier over the unlabeled data and the known fraction of positives which is minimized when these terms are equal. This approach is an instance of a general class of posterior regularization called generalized expectation (GE) criteria, as specifically proposed by Mann and McCallum[19]. However, because we wish for our classifier to be a neural network and to optimize the objective using minibatched stochastic gradient descent, the gradient of the objective must be approximating using samples from the data. Estimates of the gradient of the GE-KL objective from samples are biased, which could cause SGD to find a suboptimal solution.

To address this issue, we propose an alternative GE criteria, GE-binomial, defined so as to minimize the difference between the distribution over the number of positives in the minibatch and the binomial distribution parameterized by $\pi$. The number of positive data points, k, in a minibatch of N samples from U follows the binomial distribution with parameter $\pi$. Furthermore, the classifier g also describes a distribution over the number of positives in the minibatch as

$$q(k) = \sum_{y \in Y(k)} \prod_{i=1}^{N} g(x_i)^{y_i}\left(1 - g(x_i)^{(1-y_i)}\right)$$

where x is a micrograph region, y is an indicator vector ($y_i \in \{0,1\}$) denoting which data points are positive ($y_i = 1$) and negative ($y_i = 0$) and Y(k) is the set of all such vectors summing to k.

This allows us to define the new GE criteria as the cross entropy between these two distributions $\sum_{k=1}^{N} q(k) \log p(k)$ giving the full GE-binomial objective function

$$E_{x \sim P}[L(g(x), 1)] + \lambda \sum_{k=1}^{N} q(k) \log p(k) \qquad \text{(GE-binomial)}$$

In practice, because computing exact $q(k)$ is slow, we make a Gaussian approximation with mean $\sum_{i=1}^{N} g(x_i)$ and variance $\sum_{i=1}^{N} g(x_i)(1 - g(x_i))$ and substitute the Gaussian PDF with these parameters for $q$ in the above equation.

## 5. Autoencoder-based classifier regularization

When including the autoencoder component, we break our classifier network into two components: an encoder network composed of all layers except the final linear layer and the linear classifier layer. We denote these networks as $f$ and $c$, respectively, with the full network, $g$, being given by $g(x) = c(f(x))$. Furthermore, we introduce a deconvolutional (also called transposed convolutional, see next section) decoder network, $d$, which takes the output of the feature extractor network and returns a reconstruction of the input image, $x' = d(f(x))$. The objective function is then modified to include a term penalizing the expected reconstruction error over all images in the dataset, $D$, with weight $\gamma$

$$E_{x \sim P}[L(c(f(x)), 1)] + \lambda \sum_{k=1}^{N} q(k) \log p(k) + \gamma E_{x \sim D}\left[\|x - d(f(x))\|_2^2\right]$$

This forms the full GE-binomial objective function with autoencoder component used in Topaz.

## 6. Classifier and autoencoder architectures and hyperparameters

We use a simple three-layer convolutional neural network with striding, batch normalization[22], and parametric rectified linear units (PReLU) as the classifier in this work. The model is organized as 32 conv7x7 filters with batch normalization and PReLU, stride by 2, 64 conv5x5 filters with batch normalization and PReLU, stride by 2, 128 conv5x5 filters with batch normalization and PReLU, and a final fully connected layer with a single output.

When augmenting with an autoencoder, we use a decoder structure similar to that of DCGAN[23] . The d-dimensional representation output by the final convolutional layer of the classifier network is projected to a small spatial dimension but large feature dimension representation. This is repeatedly projected into larger spatial dimension and smaller feature dimension representations until the final output is of the original input image size. Specifically, this model is structured as repeated transpose convolutions with batch normalization and leaky ReLU activations. Let z be the representation output by the final convolutional layer of the classifier and X' be the image reconstruction given by the decoder, the decoder structure is z -> transpose conv4x4 128-d, batch normalization, leaky ReLU -> transpose conv4x4 64-d, stride 2, batch normalization, leaky ReLU -> transpose conv4x4 32-d, stride 2, batch normalization, leaky ReLU -> transpose conv3x3 1-d, stride 2 -> X'.

## 7. PU learning benchmarking

To compare classifiers trained with the different objective functions, we simulate hand labeling with various amounts of effort by randomly sampling varying numbers of particles from the training sets to treat as the positive examples. All other particles are considered unlabeled. We use cross entropy loss for the labeled particles. The values of $\pi$ used for training are specified in table 1. For GE-KL we set the GE criteria weight, $\lambda$, to 10 as recommended by Mann and McCallum[19]. For GE-binomial, we set this parameter to 1. The classifier is then trained with those positives and evaluated by average-precision score (see next section for description of classifier evaluation) on the test set micrographs. This is repeated with 10 independent samples of particles for each number of positives.

## 8. Classifier evaluation

Classifiers were evaluating by average-precision score. This score is a measure of how well ranked the micrograph regions were when ordered by the predicted probability of containing a particle and corresponds to the area under the precision-recall curve. It is calculated as the sum over the ranked micrograph regions of the precision at $k$ elements times the change in recall

$$\sum_{k=1}^{n} Pr(k)\left(R(k) - R(k-1)\right)$$

where precision (Pr) is the fraction of predictions that are correct and recall (Re) is the fraction of labeled particles that are retrieved in the top $k$ predictions

$$TP(k) = \sum_{i=1}^{k} \Sigma y_i$$

$$Pr(k) = TP(k)/k$$

$$R(k) = TP(k) \bigg/ \sum_{i=1}^{n} y_i$$

This measure is commonly used in information retrieval.

## 9. Non-maximum suppression algorithm for extracting particle coordinates

Non-maximum suppression chooses coordinates and their corresponding predicted probabilities of being a particle greedily starting from the highest scoring region. In order to prevent nearby pixels from also being considered particle candidates, all pixels within a second user-defined radius are excluded when a coordinate is selected. We set this radius to be the half major-axis length of the particle, however, smaller radii may give better results for closely packed, irregularly shaped particles.

**10. Micrograph pre-processing**

For EMPIAR-10025 and -10096[24,25], the aligned and summed micrographs along with CTF estimates were taken directly from the public data release on EMPIAR. For EMPIAR-10028[26], frames were aligned and summed without dose compensation using MotionCor2. Whole micrograph CTF estimates provided with the public release were used for this dataset.

For the clustered protocadherin dataset (Shapiro-lab), single particle micrographs were collected on a Titan Krios (Thermo Fisher Scientific) equipped with a K2 counting camera (Gatan, Inc.); the microscope was operated at 300 kV with a calibrated pixel size of 1.061 Å. 10 secs exposures were collected (40 frames/micrograph), for a total dose of 68 e$^-$/Å$^2$ with a defocus range of 1 to 4 μm. A total of 896 micrographs were collected using Leginon[27]. Frames were aligned using MotionCor2[28]. 1,540 particles were picked manually using Appion Manual Picker[20] from 87 micrographs and used as a training dataset for Topaz.

The rabbit muscle aldolase dataset (NYSBC-aldo) was collected on a Titan Krios (Thermo Fisher Scientific) equipped with a K2 counting camera (Gatan, Inc.) in super-resolution mode; the microscope was operated at 300 kV with a calibrated super-resolution pixel size of 0.416 Å. 6 secs exposures were collected (30 frames/micrograph), for a total dose of 70.32 e$^-$/Å$^2$ with a defocus range of 1 to 2 μm. A total of 1,052 micrographs were collected using Leginon[27]. Frames were aligned, Fourier binned by a factor of 2, and dose compensated using MotionCor2[28]. Whole-image CTF estimation was performed using CTFFIND4[29].

## 11. 3D reconstruction procedure

Reconstruction was performed using cryoSPARC[30]. For each particle set, we first generated an ab initio structure with a single class. These structures were then refined using cryoSPARC's "homogenous refinement" option with symmetry specified depending on the dataset (T20S proteasome: D7, 80S ribosome: C1, aldolase: D2). For the aldolase dataset, we used C2 symmetry for ab initio structure determination. Otherwise, all other parameters were left in the default setting. When evaluating the quality of Topaz particle sets for decreasing score thresholds, each particle set was selected by taking all particle predicted by the Topaz model with score greater than or equal to the given threshold. Reconstructions were calculated for each of these sets independently as described above.

## 12. 2D class averages

Class averages were calculated using the cryoSPARC "2D Classification" option. All settings were left as default except the number of 2D classes which was set to 10 for every particle set.

## 13. 3D structure analysis

The final 3D reconstructions were analyzed visually in UCSF Chimera[31] and with 3DFSC[25]. In Chimera, the published/previous 3D reconstruction was first loaded (with the fit PDB structure, if available) to which the newly-processed 3D reconstruction was then aligned. The structures were visually compared and representative areas were chosen for display in figure 4. The 3DFSCs were calculated using the public server, https://3dfsc.salk.edu, which compares Fourier shell components for several solid angles to determine the range of resolutions and the amount of anisotropy in the reconstruction.

## Code availability statement

Source code for Topaz is publicly available on GitHub at https://github.com/tbepler/topaz. Topaz is licensed under the GNU General Public License v3.0.

## Data availability statement

Single particle half maps, full sharpened maps, and masks for XXXX have been deposited to the Electron Microscopy Data Bank (EMDB) with accession codes EMD-XXXX. The full rabbit muscle aldolase dataset (NYSBC-aldo) has been deposited to the Electron Microscopy Pilot Image Archive (EMPIAR) with accession code EMPIAR-XXXX.

## Acknowledgements

The authors wish to thank Simons Electron Microscopy Center (SEMC) OPs for the aldolase sample preparation and collection, Yong Zi Tan (Columbia University) for SPA discussion, and the Electron Microscopy Group at the New York Structural Biology Center (NYSBC) for microscope calibration and assistance. We would also like to thank Tommi Jaakkola (MIT) for his valuable feedback on the machine learning methods.

T.B., A.M., and B.B. were supported by NIH grant R01-GM081871. A.J.N. was supported by a grant from the NIH National Institute of General Medical Sciences (NIGMS) (F32GM128303). This work was performed at the SEMC and National Resource for Automated Molecular Microscopy located at NYSBC, supported by grants from the Simons Foundation (SF349247), NYSTAR, and the NIH NIGMS (GM103310) with additional support from the Agouron Institute (F00316) and NIH (OD019994).

## Author contributions

T.B., A.M., and B.B. conceived of this project. T.B. developed GE-binomial and implemented Topaz, processed and analyzed single particle datasets, and carried out the computational experiments, under the guidance of B.B. J.B. prepared and collected the clustered protocadherin dataset. A.J.N. analyzed the single particle cryoEM reconstructions. T.B., A.M., J.B., L.S., A.J.N., and B.B. designed the experiments. T.B., A.J.N., and B.B. wrote the manuscript.

## Competing financial interests

The authors declare no competing financial interests.

# Tables

### Table 1 (Dataset summary)

| Dataset | Protein | Original (ang/pix) | Downsampled (ang/pix) | Particle size (pix) | Training radius (pix) | $\pi$ | Train | | Test | |
|---|---|---|---|---|---|---|---|---|---|---|
| | | | | | | | Number of micrographs | Number of particles | Number of micrographs | Number of particles |
| EMPIAR-10025 | T20S proteasome | 0.98 | 15.7 | 7 | 3 | 0.035 | 156 | 39653 | 40 | 10301 |
| EMPIAR-10028 | 80S ribosome | 1.34 | 10.7 | 12 | 3 | 0.012 | 831 | 80701 | 250 | 24546 |
| EMPIAR-10096 | Hemagglutinin trimer | 1.31 | 5.24 | 10 | 4 | 0.035 | 347 | 100465 | 100 | 29535 |
| NYSBC-aldo | Rabbit muscle aldolase | 0.83 | 6.64 | 10 | 3 | 0.1 | 865 | 163758 | 200 | 39347 |
| Shapiro-lab | Clustered protocadherin | 1.061 | 8.49 | 15 | 4 | 0.015 | 67 | 1167 | 20 | 373 |

**Table 1** | Summary of cryoEM datasets and hyperparameters used for classifier training on each. Each dataset was downsampled and split into train and test sets at the whole micrograph level.

# Figures

**Figure 1 (Single particle cryoEM workflow)**

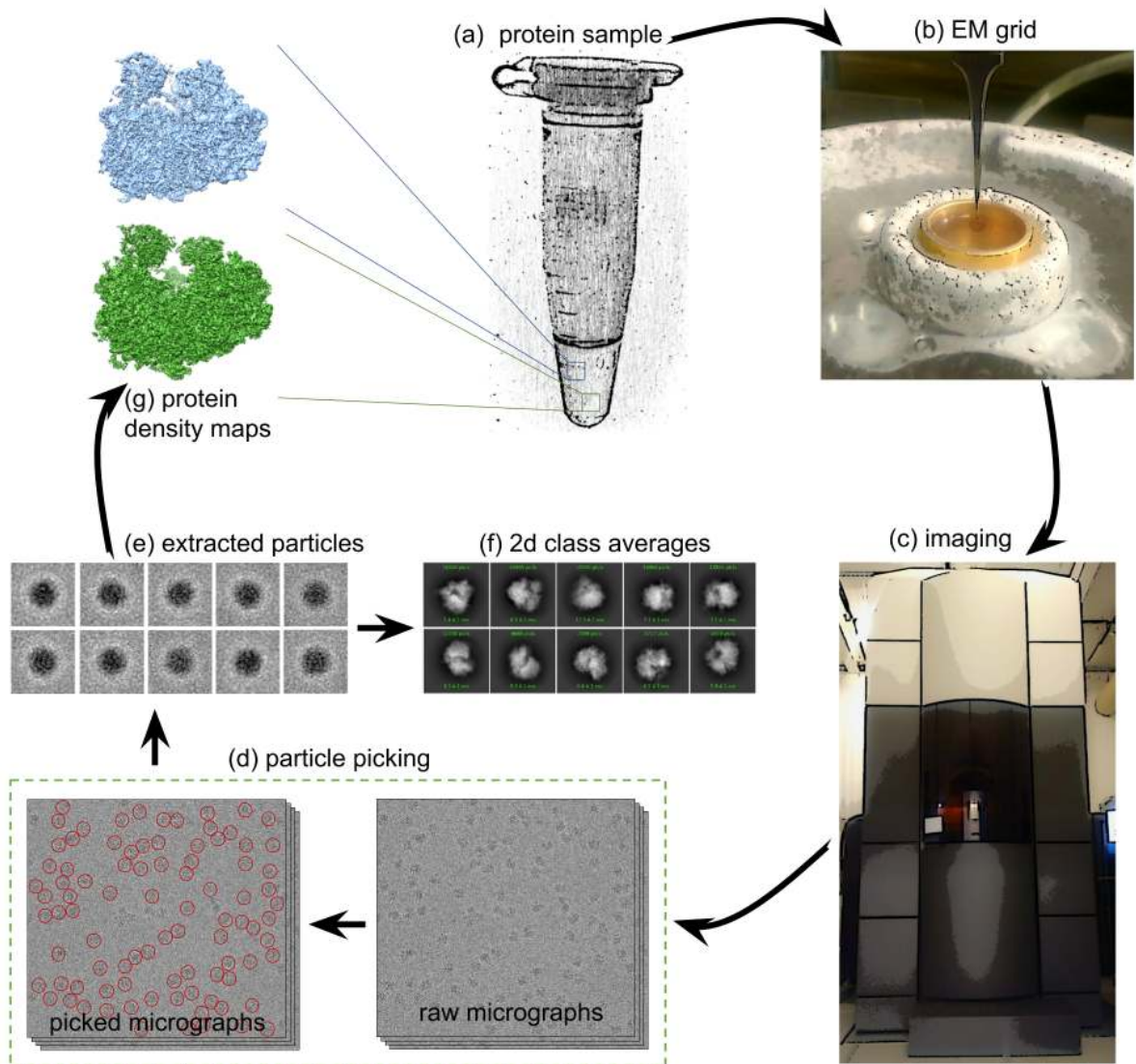

**Figure 1** | Single particle cryoEM entails spreading a thin (typically <100 nm) layer of purified protein sample **(a)** across an EM grid, vitrifying the sample **(b)**, collecting hundreds or thousands of TEM micrographs **(c)**, picking a sufficient amount of real particles **(d)** and extracting the particle stack **(e)**, optionally performing 2D classification **(f)** (contrast inverted), and aligning those particles in 3D space to produce potentially multiple different electron density maps **(g)**. Topaz focuses on optimizing the particle picking step boxed in green.

**Figure 2 (Topaz pipeline diagram)**

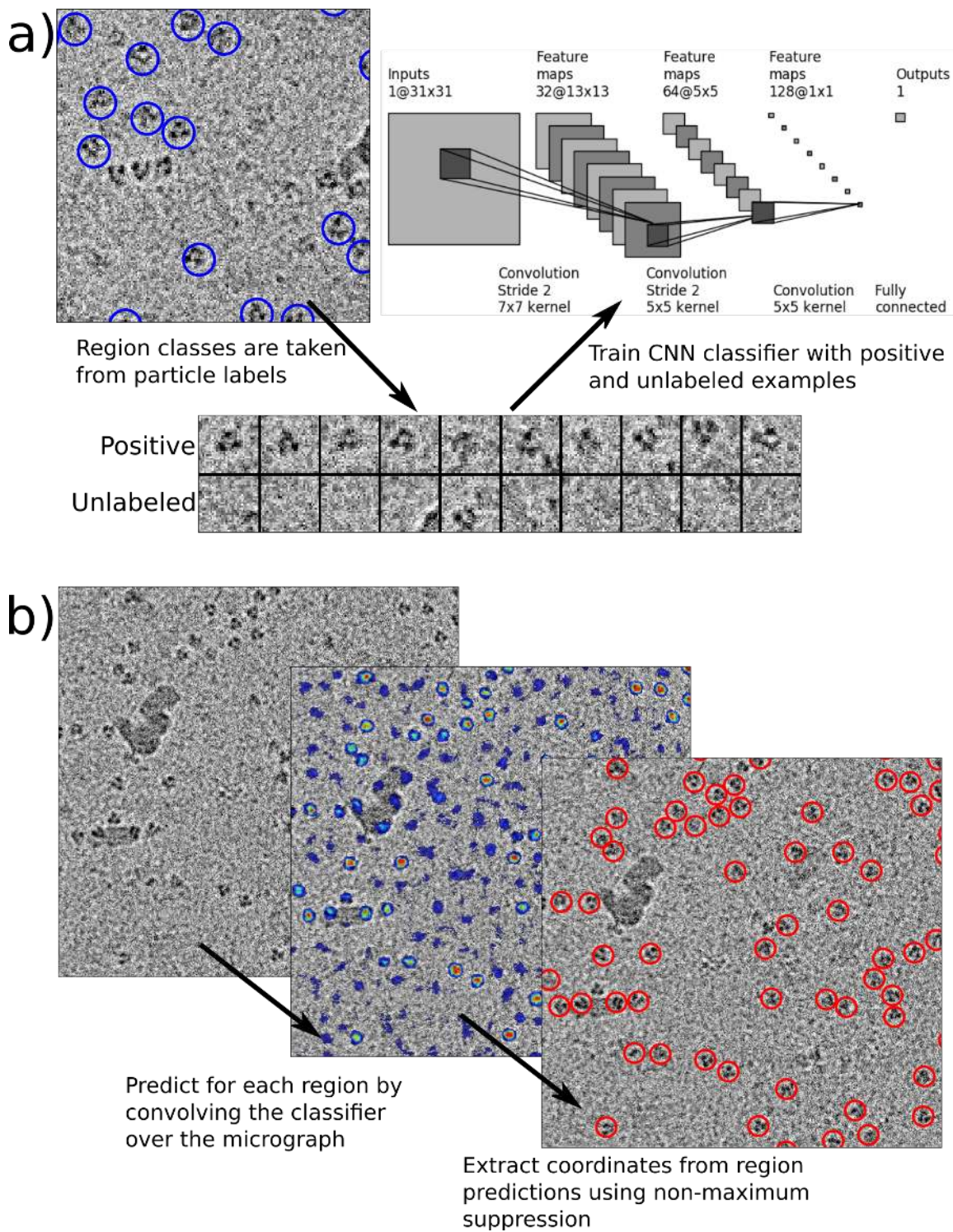

**Figure 2** | Topaz particle picking pipeline using CNNs trained with positive and unlabeled data. (**a**) Given a set of labeled particles, a CNN is trained to classify positive and negative regions using particle locations as positive regions and all other regions as unlabeled. Labeled particles

from EMPIAR-10096 are indicated by blue circles and a few positive and unlabeled regions are depicted. **(b)** Once the CNN classifier is trained, particles are predicted in two steps. First, the classifier is convolved over each micrograph to give predictions for each region of each micrograph. Then, coordinates are extracted from the region predictions using non-maximum suppression. The left image shows a raw micrograph from EMPIAR-10096. The middle image depicts the micrograph with overlayed region predictions [blue = low confidence, red = high confidence]. The right image indicates predicted particles after using non-maximum suppression on the region predictions.

**Figure 3 (Reconstructions)**

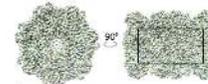
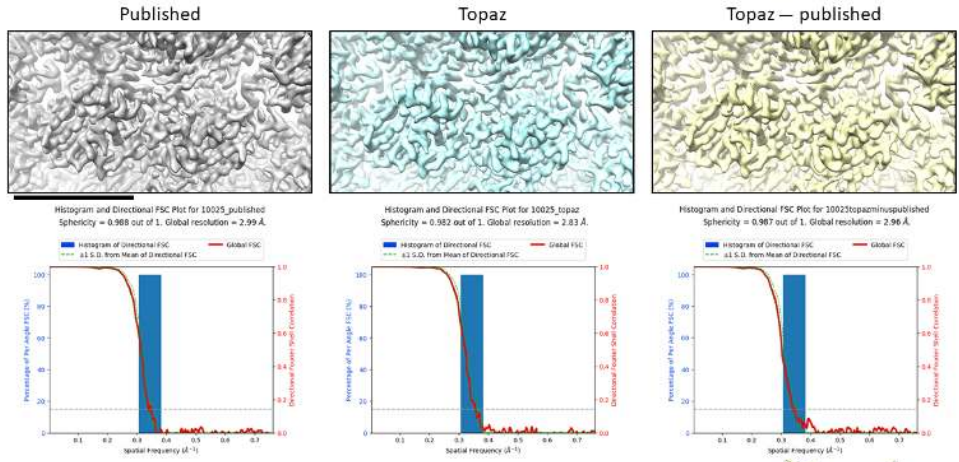
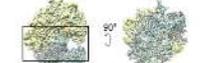
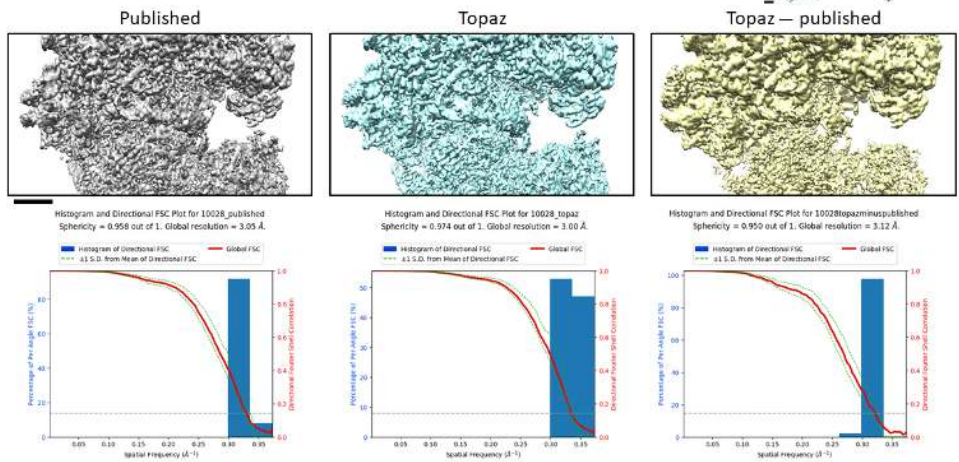
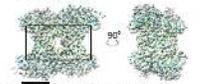
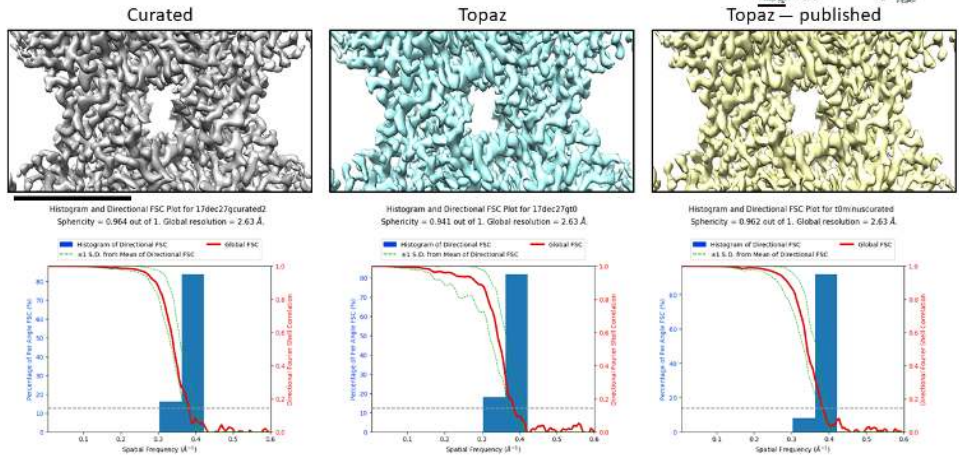

**Figure 3** | Single particle reconstructions from published particles, Topaz particles, and Topaz particles with published particles removed (left to right) showing that the additional particles picked by Topaz result in nearly identical maps. **(a)** T20S proteasome (EMPIAR-10025) using the provided aligned, dose-weighted micrographs. Topaz picked 3.22x more particles than the published picks, resulting in a 0.16 Å increase in resolution. **(b)** 80S ribosome (EMPIAR-10028). Topaz picked 1.72x more particles than the published picks, resulting in the highest resolution structure of this dataset to date. **(c)** Rabbit muscle aldolase. Topaz picked 3.68x more particles than the published picks, however the resolution did not improve. This suggests that particle number is not the resolution-limiting factor with this dataset; per-particle CTF, per-particle frame alignment, and other potential factors such as beam tilt correction may be limiting the resolution. Scale bars: 3 nm

**Figure 4 (Ranked particles vs. resolution, 2d class false positives)**

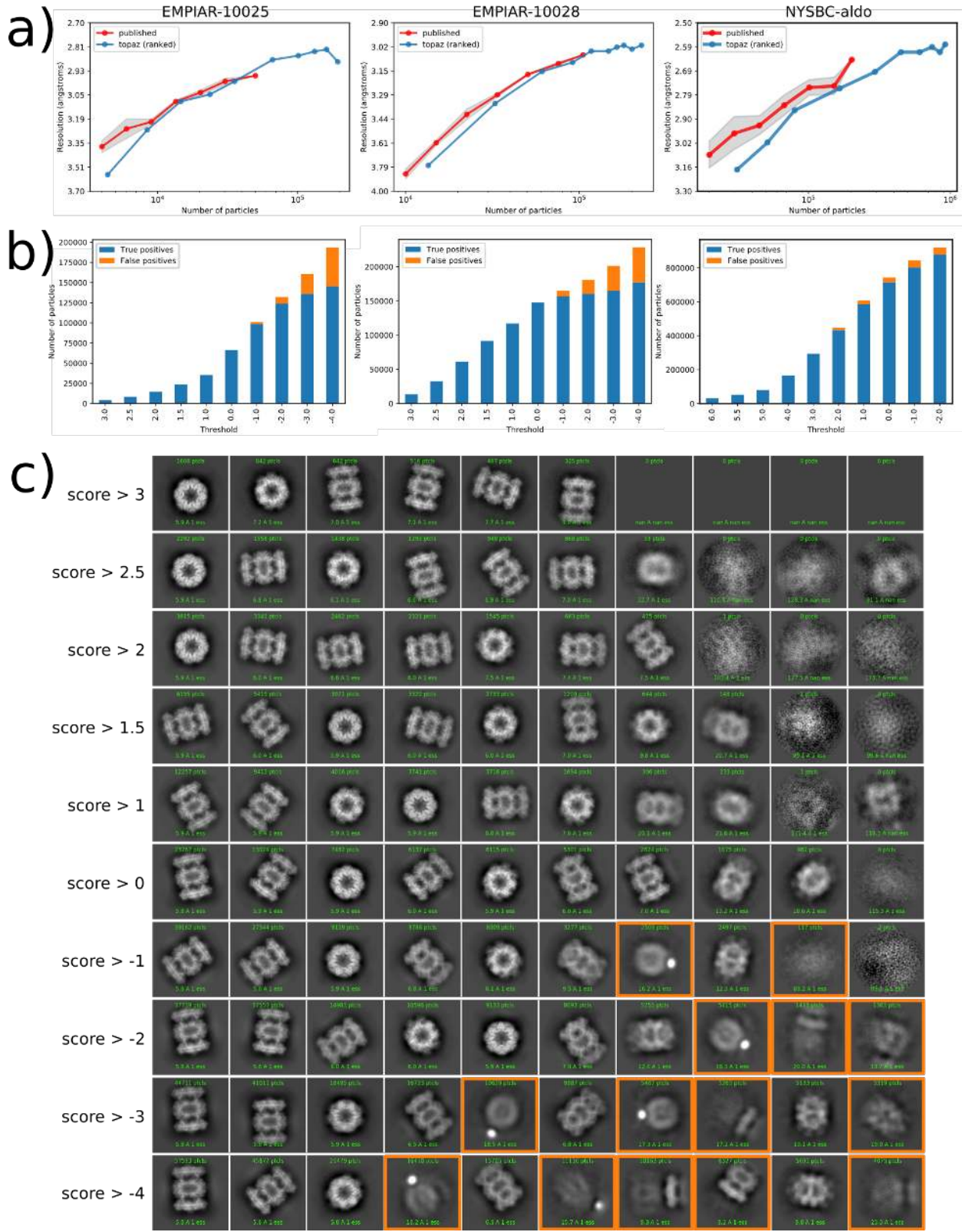

**Figure 4** | Reconstruction resolution and 2d class averages for Topaz particles at decreasing log-likelihood ratio thresholds. **(a)** Number of particles vs. reconstruction resolution for Topaz particles (increasing number of particles corresponds to decreasing log-likelihood threshold) and randomly sampled subsets of the published particle set. Resolution is as reported by cryoSPARC. For the published particle sets the mean of three replicates is marked with standard deviation shaded in grey. **(b)** Stacked bar plots show the quantification of the number of true and false positives at each threshold based on 2d class averages. Decreasing threshold corresponds to increasing number of predicted particles. True positives are colored in blue and false positives in orange. **(c)** 2d class averages obtained at each score threshold for the T20S proteasome (EMPIAR-10025). Classes determined to be false positives are marked with orange boxes. We note that several classes which appear to be false positives at high score thresholds do not contain any particles and, therefore, are not highlighted.

**Figure 5 (Classifier comparison)**

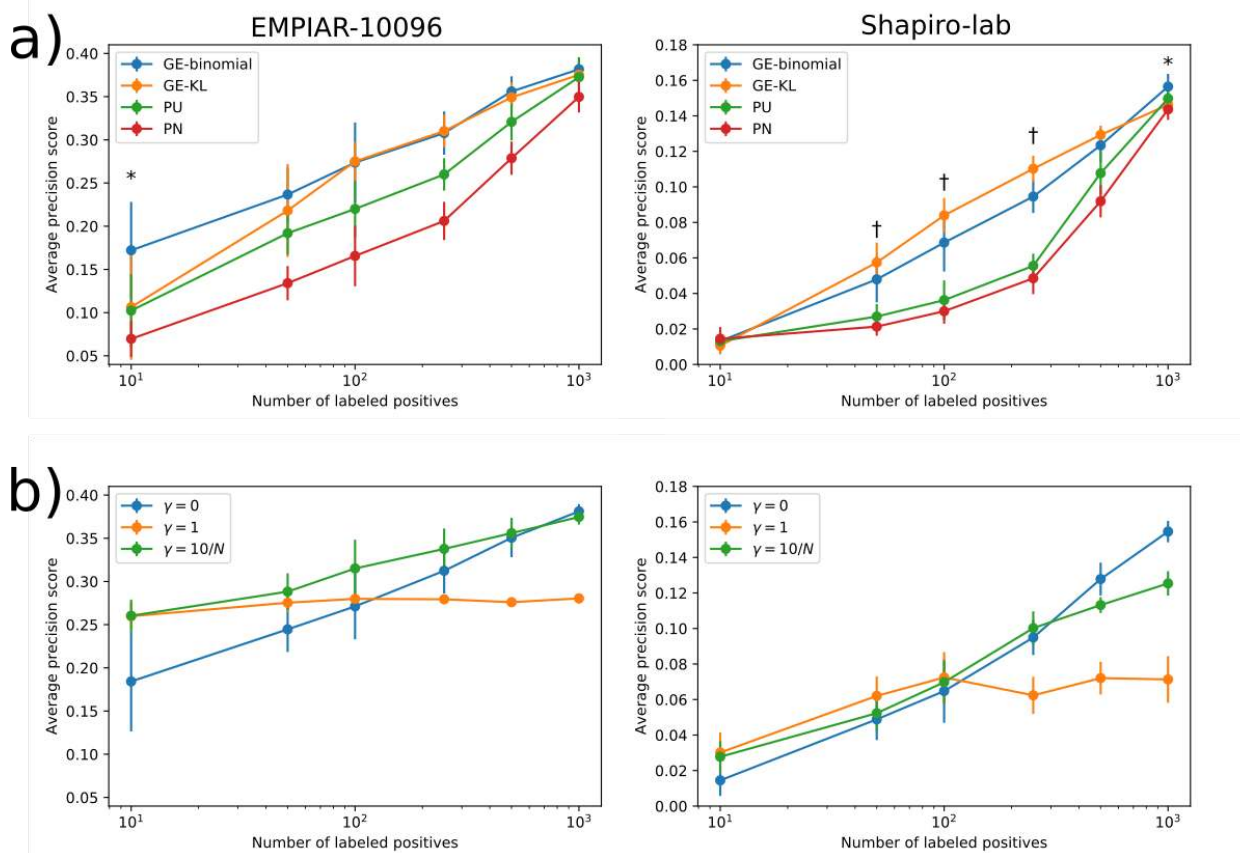

**Figure 5** | Comparison of models trained using different objective functions with varying numbers of labeled positives on the EMPIAR-10096 and Shapiro-lab datasets. **(a)** Plots show the mean and standard deviation of the average-precision score for predicting positive regions in the EMPIAR-10096 and Shapiro-lab test set micrographs for models trained using either the naive PN, Kiryo et al.'s non-negative risk estimator (PU), GE-KL, or GE-binomial objective function. (*) indicates experiments in which GE-binomial achieved higher average-precision than GE-KL with $p < 0.05$. (†) indicates experiments in which GE-KL achieved higher average-precision than GE-binomial with $p < 0.05$. **(b)** Plots showing the mean and standard deviation of the average-precision score for models trained jointly as autoencoders with different reconstruction loss weights ($\gamma$). $\gamma=0$ corresponds to training the classifier without the autoencoder. $\gamma=10/N$ means the reconstruction loss is weighted by 10 divided by the number of labeled positives used to train the model.

# Supplemental Figures

**Supplemental Figure 1 (EMPIAR-10096 example micrographs)**

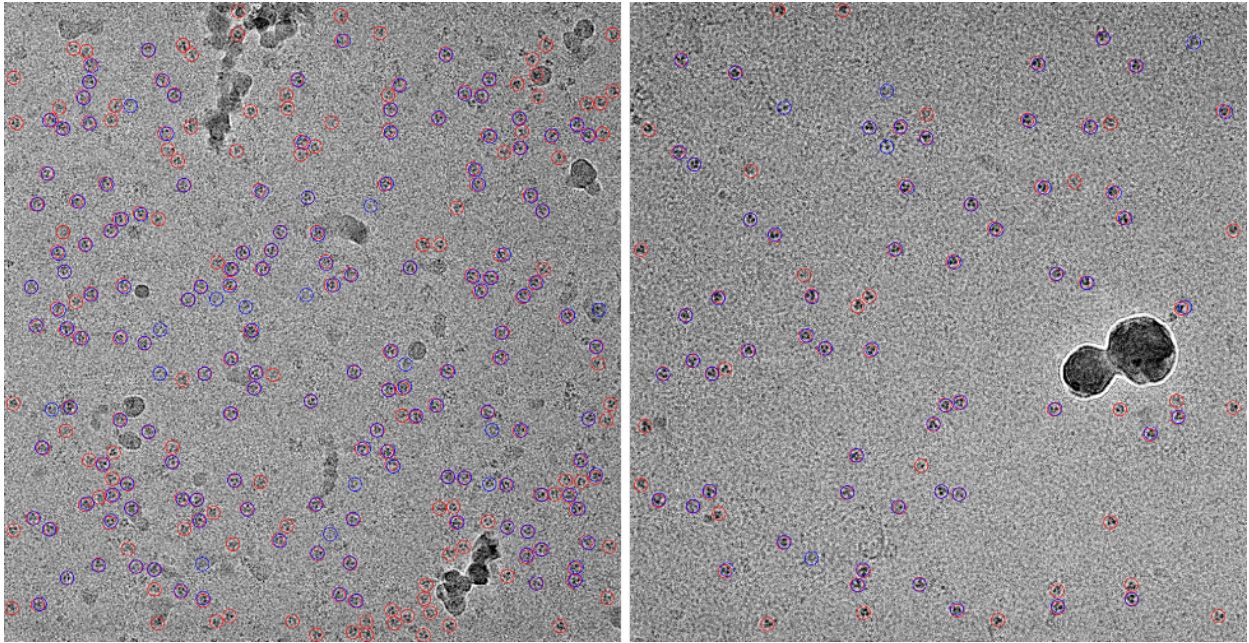

Representative micrographs from the EMPIAR-10096 test set. Ground truth (blue) and predicted (red) particles are circled. Topaz avoids ice chunks and retrieves many particles not included in the ground truth set.

**Supplemental Figure 2 (Shapiro-lab example micrograph)**

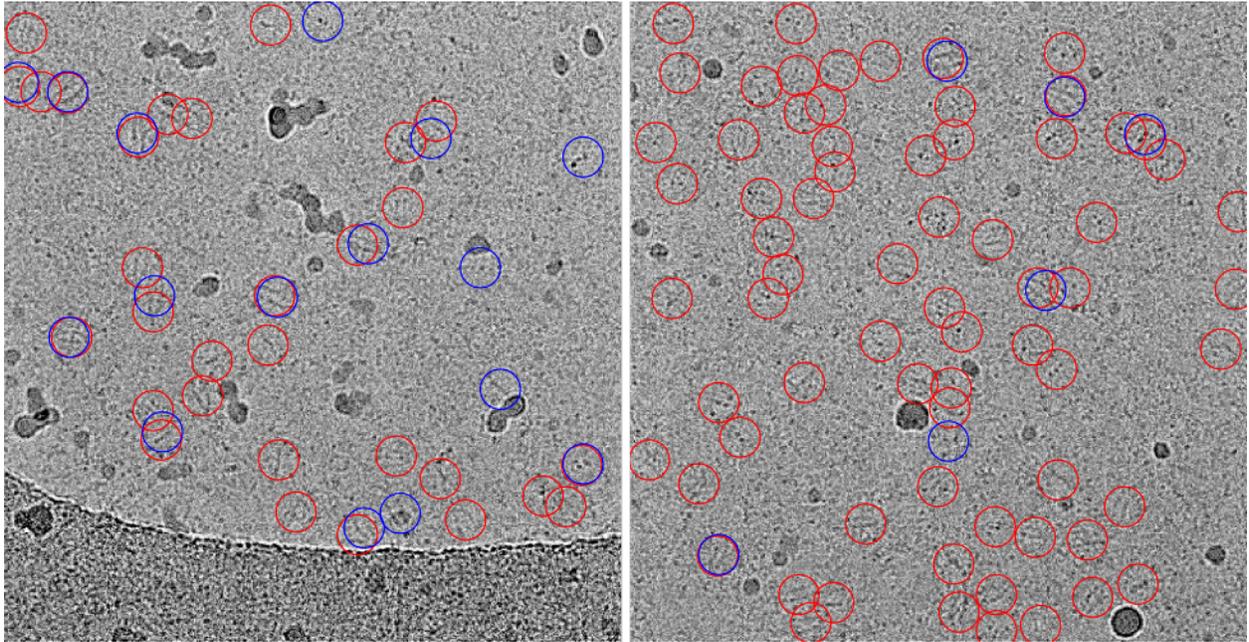

Representative micrographs from the Shapiro-lab test set. Manually selected (blue) and predicted (red) particles are circled. This is an extremely challenging dataset of a ~75kDa stick-lick protein with low signal-to-noise ratio. Topaz successfully avoids ice chunks, particles in proximity to the edge of the hole, and particles on carbon and correctly identifies many particles missing from the manually labeled set.

**Supplemental Figure 3 (Sensitivity to π of GE-binomial)**

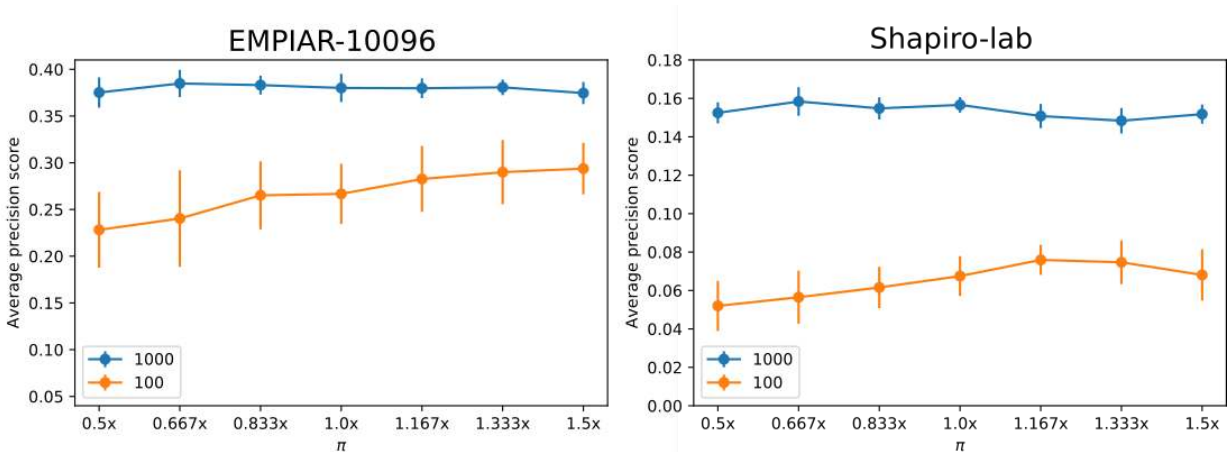

Sensitivity of GE-binomial objective function to the setting of **π**. For EMPIAR-10096 and the Shapiro-lab datasets we report average-precision scores for classifiers trained with 100 and 1000 labeled particles and values of **π** varying from 0.5x to 1.5x the values reported in table 1. We report the mean and standard deviation of 10 runs.

**Supplemental Figure 4 (EMPIAR-10028 2d class averages)**

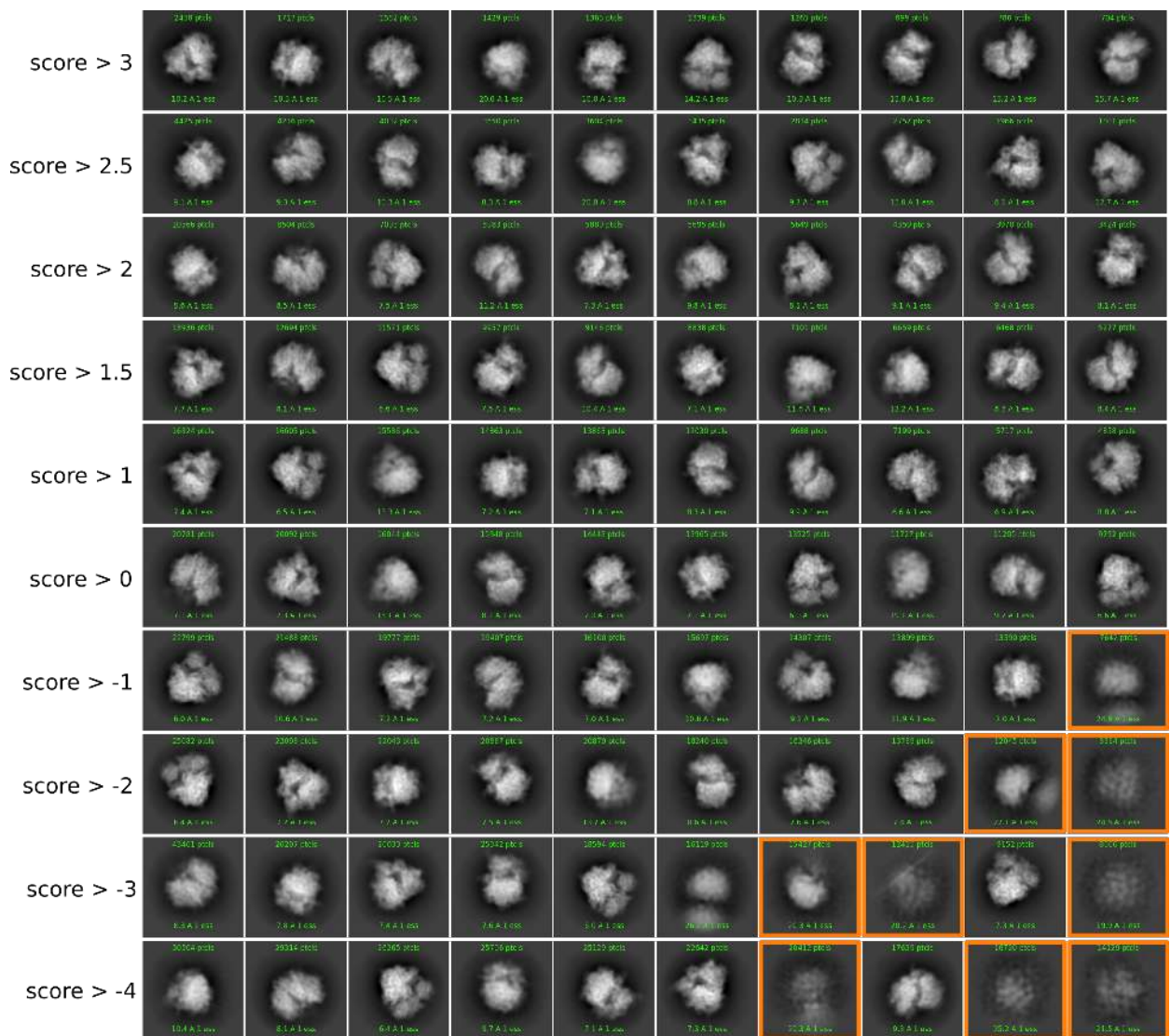

2d class averages of Topaz particles with decreasing score threshold for the 80S ribosome. Classes identified as false positives for quantification in Figure 5 are indicated by orange boxes.

**Supplemental Figure 5 (NYSBC-aldo 2d class averages)**

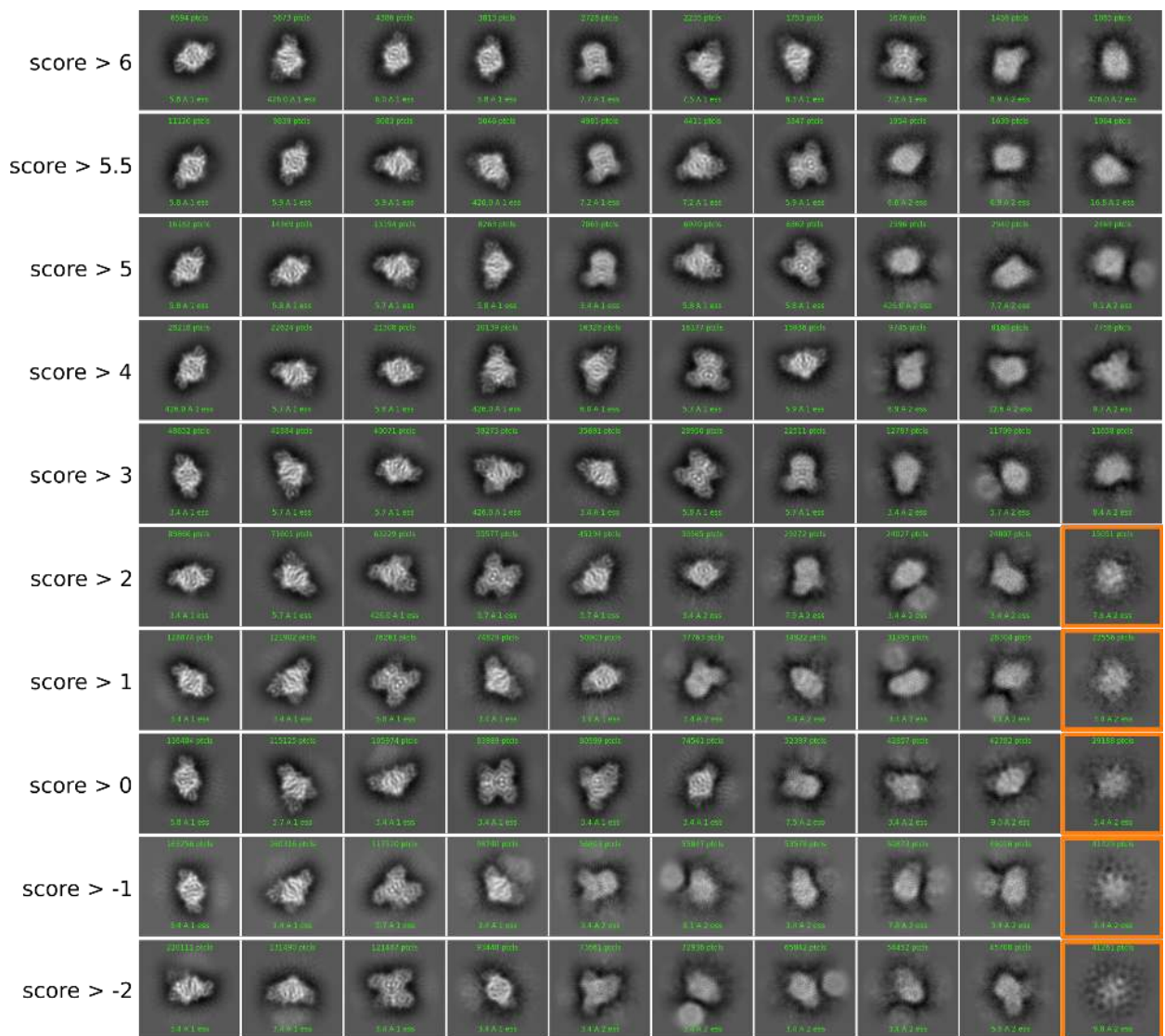

2d class averages of Topaz particles with decreasing score threshold for the aldolase dataset. Classes identified as false positives for quantification in Figure 5 are indicated by orange boxes.

**Supplemental Figure 6 (Precision-recall / F1 curves for EMPIAR-10025, EMPIAR-10028, and NYSBC-aldo)**

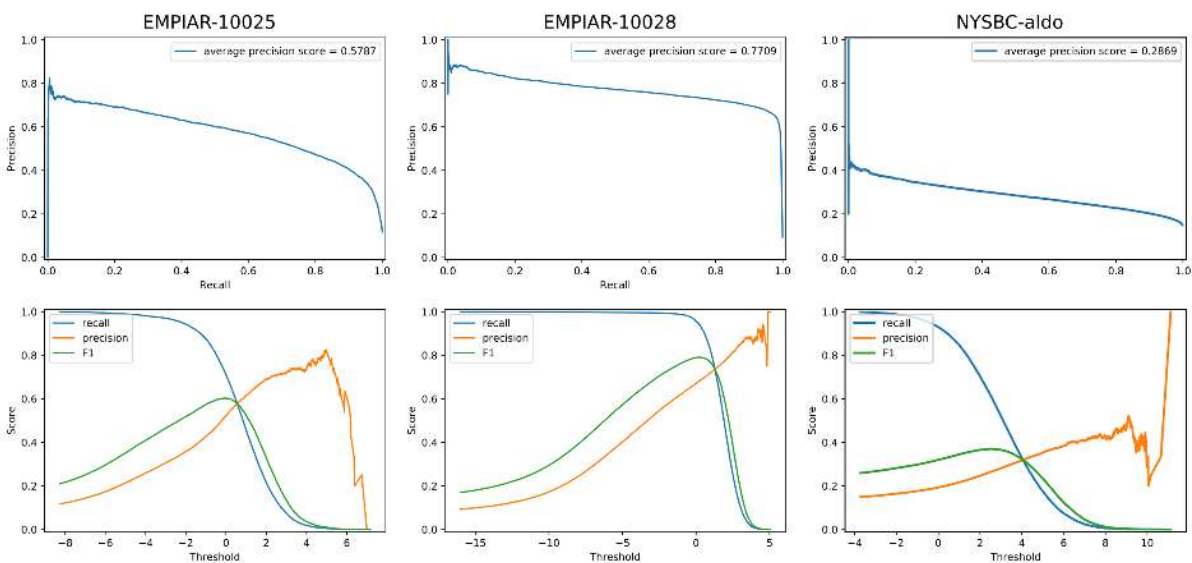

Precision-recall curves and threshold vs precision, recall, and F1 score curves for classifiers trained on the EMPIAR-10025, -10028, and NYSBC-aldo datasets. Curves were calculated by matching the particles predicted by the Topaz models on the test set micrographs of each dataset to the published particle annotations on those micrographs.